\documentclass[amsmath, amssymb,aps]{revtex4-2}
\usepackage{graphicx}
\usepackage{geometry}
\usepackage[margin=2mm]{caption}
 
\usepackage{fancyhdr}
\usepackage{lastpage}
\usepackage{url}
\usepackage{xcolor}

\usepackage[page,toc,titletoc,title]{appendix}

\usepackage{algorithm}
\usepackage{algpseudocode}
\algblock{Input}{EndInput}
\algnotext{EndInput}
\algblock{Output}{EndOutput}
\algnotext{EndOutput}
\newcommand{\Desc}[2]{\State \makebox[11em][l]{#1}#2}

\usepackage[version=4]{mhchem}

\usepackage{amssymb}
\usepackage{amsmath}
\usepackage{bm}
\usepackage{siunitx}

\numberwithin{equation}{section}

\newcommand{\ie}{\textit{i.e.} }

\newcommand{\eg}{\textit{e.g.} }
\newcommand{\Eg}{\textit{E.g.} }

\newcommand{\shh}{V_{H}}
\newcommand{\SHH}{$\shh$ }
\newcommand{\dshh}{\left|\Delta\shh\right|}
\newcommand{\DSHH}{$\dshh$ }

\newcommand{\Secref}[1]{Section \ref{#1}}
\newcommand{\Appref}[1]{Appendix \ref{#1}}
\newcommand{\Algorref}[1]{Algorithm \ref{#1}}
\newcommand{\Figref}[1]{Figure \ref{#1}}

\newcommand{\OO}[1]{\mathcal{O}\left({#1}\right)}
\newcommand{\RR}{\bm{\mathbb{R}}}

\usepackage{amsthm}
\theoremstyle{definition}

\usepackage{float}
\usepackage{placeins}
\usepackage{subcaption}

\begin{document}

\title{A Genetic Algorithm For Convex Hull Optimisation}
\author{Scott Donaldson, Robert A. Lawrence, Matt I. J. Probert.}
\email{matt.probert@york.ac.uk} 
\affiliation{School of Physics, Engineering and Technology, University of York, Heslington, North Yorkshire, YO10 5DD, 
 UK}
\date{\today}

\newcommand{\orcidauthorA}{0000-0002-0102-3162} 
\newcommand{\orcidauthorB}{0000-0003-0025-2497} 
\newcommand{\orcidauthorC}{0000-0002-1130-9316} 


\begin{abstract}
{Computationally efficient and automated generation of convex hulls is desirable for high throughput materials discovery of thermodynamically stable multi-species crystal structures. A convex hull genetic algorithm is proposed that uses methodology adapted from multi-objective optimisation techniques to optimise the convex hull itself as an object, enabling efficient discovery of convex hulls for $N \geq 2$ species. This method, when tested on a LiSi system utilising pre-trained machine learned potentials, was found to be able to efficiently discover reported structures as well as new potential LiSi candidate structures.}
\end{abstract}

\maketitle


\section{Introduction}

\textit{Convex hulls} (CHs) are a powerful tool to enable the prediction of thermodynamically stable compounds of 2 or more elements. Through proper interpretation, they can be used to predict optimal starting ratios of different elements in order to form stable products, as well as how sensitive the given ratio is towards forming a different system. This can be of great use for guiding experimental synthesis of given compounds by helping to avoid disproportionation and growth of other phases within the sample.


In order to construct a CH, the complete set of stable structures at a given temperature and pressure must be known. These structures lie at minima of the \textit{Potential Energy Surface} (PES) in the high dimensional crystal structure parameter space. Unfortunately, the full PES is not practically attainable for all but the most simple cases. Further, a search for only the minima can be computationally expensive, due in part to the number of local minima in the PES increasing exponentially with system size \cite{pickard_ab_2011}.

In recent years multiple structural databases have emerged which combine literature reported structures as well as high throughput \textit{density functional theory} (DFT) calculations to allow a user to plot a CH from known structures with trivial effort \cite{saal_materials_2013,kirklin_open_2015,jain_commentary_2013}. However, thoroughly explored searches are not present for all combinations of elements. This is especially true away from room temperature and pressure (RTP) where the paucity of experiments leads to very few hulls being complete. To counteract this, efficient computational \textit{Crystal Structure Prediction} (CSP) methods to generate CHs are desirable, especially automated methods which do not rely on the quality of the intuition of its user.

Many algorithms have been developed for CSP including (but not limited to) Monte-Carlo like (with or without constraints) ab-initio random structure searching (AIRSS) \cite{pickard_ab_2011}, simulated annealing \cite{doll_structure_2008}, basin hopping \cite{burnham_crystal_2019}, particle swarm optimisation \cite{wang_calypso_2012} and \textit{genetic algorithms} (GAs). The CASTEP GA \cite{abraham_periodic_2006,abraham_improved_2008,higgins_simultaneous_2019} is used as the basis for this paper, though other CSP GA implementations exist \cite{curtis_gator_2018,lonie_xtalopt_2011,oganov_crystal_2006,GASP-Python}.

The results for many CSP methods have been successfully used to find convex hulls for a variety of systems \cite{ferreira_search_2023,simon_magnetism_2018,nguyen-cong_tinselenium_2018,tsuppayakorn-aek_phonon-mediated_2023,wang_prediction_2021}. However, CSP methods commonly do not optimise the convex hull as an object. A common approach is to optimise towards the global minima in the PES with some penalisation applied to encourage a spread of solutions. This allows non-global minima found `on the way' to be used to construct the CH `for free'. This paper proposes a modification to existing genetic methods to optimise the CH directly by borrowing from multi objective optimisation methods to create the CASTEP \textit{Convex Hull Genetic Algorithm} (CHGA).

An outline of CSP GAs, with specific focus on the CASTEP GA, is given in Section \Secref{sec:intro_to_GAs} (more detail on the CASTEP GA can be found in \cite{abraham_periodic_2006,abraham_improved_2008,higgins_simultaneous_2019}). As the CHGA builds on the CASTEP GA \Secref{sec:convex_hull_ga} focuses on the changes made to the CASTEP GA to create the CHGA. The methods used for testing the CHGA are outlined in \Secref{sec:testing_methods_and_ML} with results of this testing on a LiSi system given in \Secref{sec:results}.


\section{Background Theory}
\label{sec:intro_to_GAs}

\textit{Genetic algorithms} (GAs) (a subset of \textit{evolutionary algorithms}) are a family of natural programming techniques originally developed in the 1970's \cite[Ch 21]{rozenberg_handbook_2012}. GAs model Darwinistic evolution by evolving a population of solutions to some problem in order to optimise towards one or more optimal solutions. This may seem an unusual approach to optimisation of a non-organic process, but GAs have shown promise in a variety of applications to CSP \cite{abraham_periodic_2006,abraham_improved_2008,curtis_gator_2018,lonie_xtalopt_2011,oganov_crystal_2006,ferreira_search_2023,simon_magnetism_2018,nguyen-cong_tinselenium_2018,tsuppayakorn-aek_phonon-mediated_2023,wang_prediction_2021}. Their success is in part due to GAs falling into the category of unsupervised learning techniques \cite[Ch 1.3.1.2]{otterlo_logic_2009}, \ie a population evolves towards optimality with only survival of the fittest guiding evolution, as it does in nature. This removes the requirement for any \textit{a priori} knowledge to restrict the search space, which may bias against unexpected novel structures.

CSP GAs utilise a Darwinistic approach to finding optimal structure(s) by considering sets (or \textit{populations}) of relaxed structures, ascribing to each \textit{member} of the population a \textit{fitness} proportional to their optimality in comparison to all other population members, where the CASTEP GA calculates fitness based on relaxed cell enthalpy \cite{abraham_improved_2008}. Pairs of \textit{parent} members are chosen based on their fitness (in the CASTEP GA \cite{abraham_periodic_2006} using roulette selection) and combined to create new \textit{child} members in a \textit{breeding} (or \textit{crossover}) process.

Breeding in the CASTEP GA utilises periodic slices across parent cells, combining the slices to create a new child (as outlined in \cite{abraham_periodic_2006}). By combining elements of both parents in such a way the distinct children inherit information from both parents. Noise is then randomly added to the children's ionic positions and/or lattice vectors as an analogue to genetic \textit{mutation}. Mutation counteracts stagnation of the population, stopping the population from `getting stuck' in one or more minima \cite{greenstein_determining_2023}.

Post-mutation children are \textit{relaxed} using a \textit{geometry optimisation} towards a local minima in the PES and the fitness of this generation of parents and children evaluated. The CASTEP GA calculates fitness based on the cell enthalpies and the structural similarity of a member to the most stable member of that stoichiometry \cite{abraham_improved_2008}. Members are then selected to survive to be the next \textit{generation} of parents. In the CASTEP GA this selection is \textit{greedy} (from the combined set of parents and children) and \textit{semi-elitist}; the best members are always taken forwards with the remainder chosen for survival with fitness based roulette selection \cite{abraham_periodic_2006}. This repeats for multiple generations, with the population evolving towards one containing the lowest enthalpy structure.


\section{The Convex Hull Genetic Algorithm}
\label{sec:convex_hull_ga}

The CASTEP \textit{convex hull genetic algorithm} (CHGA) builds on the CASTEP GA \cite{abraham_periodic_2006,abraham_improved_2008,higgins_simultaneous_2019}, modifying the methods used to calculate the fitness of population members and how to select members based on this fitness. This is done by adapting some approaches from \textit{multi-objective genetic algorithms} (MOGAs) for the specific problem of convex hulls. This section introduces all the points of difference between the CASTEP GA and the CASTEP CHGA, treating each MOGA method and how it has been adapted individually.


\subsection{Binding Enthalpy}

In a standard GA generally fitness is derived from a cell enthalpy (per ion). However, this cannot be used when comparing two structures with different stoichiometries as one would expect different stoichiometries to differ in enthalpy. For a crystal containing species $\in\{a,b,\dots\}$, the binding enthalpy of the cell is
\begin{equation}
  \label{eq:def_binding_enthalpy}
  \Delta H
  :=
  \frac{ H_{T} - \sum_{x\in\{a,b,\dots\}} N_{x}\mu_{x} }{ \sum_{x\in\{a,b,\dots\}} N_{x} }
  \quad ,
\end{equation}
where $H_{T}$ is the total cell enthalpy and $N_{x},\mu_{x}$ are the number and chemical potential of the ions of species $x\in\{a,b,\dots\}$ respectively. In the CHGA $\mu_{x}$ is either calculated from an input \textit{pure crystal cell} (the ground state single species crystal structure for a given element) or given by the user, and by construction binding enthalpies of ground state \textit{pure cells} (single species crystals) are 0. 

When considering convex hull optimisation binding enthalpy is commonly used to compare thermodynamic stability of structures. However, there are other contributions which this ignores (for example finite temperature and/or entropic contributions), such that the hull is a the 0 K hull. In general, a convex hull optimisation is high throughput so ignoring other considerations is justified, but must be borne in mind when analysing the final population.


\subsection{The Convex Hull}

Formally, a CH of $\geq N+1$ points in $\mathbb{R}^{N}$ is the smallest subset of points such that when joined by straight edges create the closed plane (or hyperplane) with the smallest surface area such that all points are either on this plane or contained within it. This is an abstract concept, especially for $N>3$, but a common method for assessing thermodynamic stability of mixed ionic species crystal stoichiometries. Finding a convex hull from a given set of points (ignoring which space they exist within) is a well studied problem in computer science, with the Quickhull algorithm \cite{barber_quickhull_1996} implemented in the CHGA due to its efficiency and applicability without modification in $>2$ dimensions.

To parameterise population members in the CHGA in terms of the convex hull a standard dimensional reduction is applied to give the \textit{reduced species ratios} $\mathbf{R}\in\RR^{N-1}$ (for $N$ species). For example, for a cell containing two species $a,b$ with $N_{a},N_{b}$ ions of each species respectively then the \textit{ionic ratio} is $R_{a}=N_{a}/(N_{a}+N_{b})$ and similarly for $b$ where $\sum_{i\in\{a,b\}} R_{i} = 1$. The two species dimensional reduction is trivially $\mathbf{R} = R_{a}$ or with an additional third species $c$ (where $\sum_{i\in\{a,b,c\}} R_{i} = 1$)
\begin{equation}
  \label{eq:3_spec_ratio_transform_def}
  \mathbf{R}
  =
  \left( \frac{2R_{b} + R_{c}}{2}, \frac{R_{c}\sqrt{3}}{2}\right)
  \quad .
\end{equation}
This allows the convex hull to be plot from $(\mathbf{R},\Delta H) \in \RR^{N}$.

The subset of members that define the CH are referred to as \textit{on the hull} and all others as \textit{above the hull}. Often it is the thermodynamically stable structures that are of interest, so members with positive binding enthalpy are discarded when considering a final CH. Also, the ground state single species crystal structures with $0$ binding enthalpy are, by construction, always on the CH.

An important subtlety is the possibility of structures being close to, but still above, the 0 K convex hull. For example, a member with binding enthalpy $\lesssim 50$ meV above the CH is roughly within room temperature $k_{B}T$ of the hull, so depending on growth conditions \textit{etc} will likely be thermally accessible. If one of these structures is the lowest binding enthalpy for its stoichiometry then it could be thermodynamically stable and of practical interest (even though it is not on the 0 K hull). It is also certainly the case that all members within any systematic error of the hull (\eg convergence of DFT parameters, geometry optimisation convergence \textit{etc}) should be considered. Due to high throughput techniques often utilised in CSP the systematic error is, for example, likely larger than with publication quality DFT calculations. The members of interest are referred to those \textit{`on or near the hull'}.


\subsubsection{The Convex Hull in Fitness Space}
\label{sec:convex_hull_in_fitness_space}

In a CSP GA the comparative fitness derived from cell enthalpy is considered. In the CASTEP GA this is calculated first from a linear scaling of the enthalpy (based on the minimum and maximum enthalpy over the whole population) to a dimensionless fitness $\mathcal{F}\in[0,1]$; a fitness of 1 is awarded to the \textit{`fittest'} (or `best'/lowest binding enthalpy member) member and 0 the `least fit' (or `worst'/highest binding enthalpy member).

If there exists more than one population member with a given ionic ratio the fitness of all but the lowest enthalpy member (for that ionic ratio) is penalised. This is done by reducing its fitness by an amount proportional to its structural similarity to the lowest enthalpy member at the given ratio. This structural fitness penalisation, outlined in detail in \cite{abraham_improved_2008}, reduces the number of structurally similar members in the fittest members of the population.

To assess the comparative optimality of all members the population is converted to a \textit{fitness space} by combining the reduced ionic ratios and scaled fitness for each member to give $(\mathbf{R},\mathcal{F}) \in \RR^{N}$. It is in this fitness space that optimality of population members is considered. Though this space is useful for fitness parameterisation convex hulls are generally considered in \textit{binding enthalpy space}. Due to the lowest enthalpy structures not having any fitness penalisation the members on the hull in fitness space are the same members on the hull in binding enthalpy space. However, due to the fitness scaling, the optimal hull in fitness space is positive, rather than negative when converted to binding enthalpy.


\subsection{Pareto \& Convex Hull Fronts}
\label{sec:pareto_hull_and_front}

GAs optimise a single parameter. In the case of CSP an enthalpy based fitness (including a structural similarity penalisation in the case of the CASTEP GA \cite{abraham_improved_2008}). However, it is often desirable to optimise multiple parameters, or \textit{objective functions}, with a MOGA. When considering multiple objectives there is generally no single `best' solution, but instead a set of equally \textit{Pareto optimal} solutions \cite{fonseca_genetic_1993}. These optimal solutions lie on the \textit{Pareto front}, where the goal of a MOGA is to find a \textit{Pareto optimal set} of members on this front (preferably evenly spaced along it).

One approach to defining a fitness based on Pareto fronts is given by the family of \textit{Non-dominated Sorting Genetic Algorithm} (\textit{NSGA}) MOGA approaches \cite{deb_fast_2002,deb_evolutionary_2014,jain_evolutionary_2014}. NSGA methods sort a population into subsets called fronts; first the Pareto optimal set (calculated from all population members) is assigned to front 1, with these members are refereed to as \textit{`on front 1'}. Front 2 is the Pareto optimal set of all members after removal of members on front 1, and so on, until all members have been assigned to a front. All members of each front are equally fit with respect to the primary (front based) fitness.

\begin{figure}[htbp]
  \centering
  \includegraphics[width=0.5\textwidth]{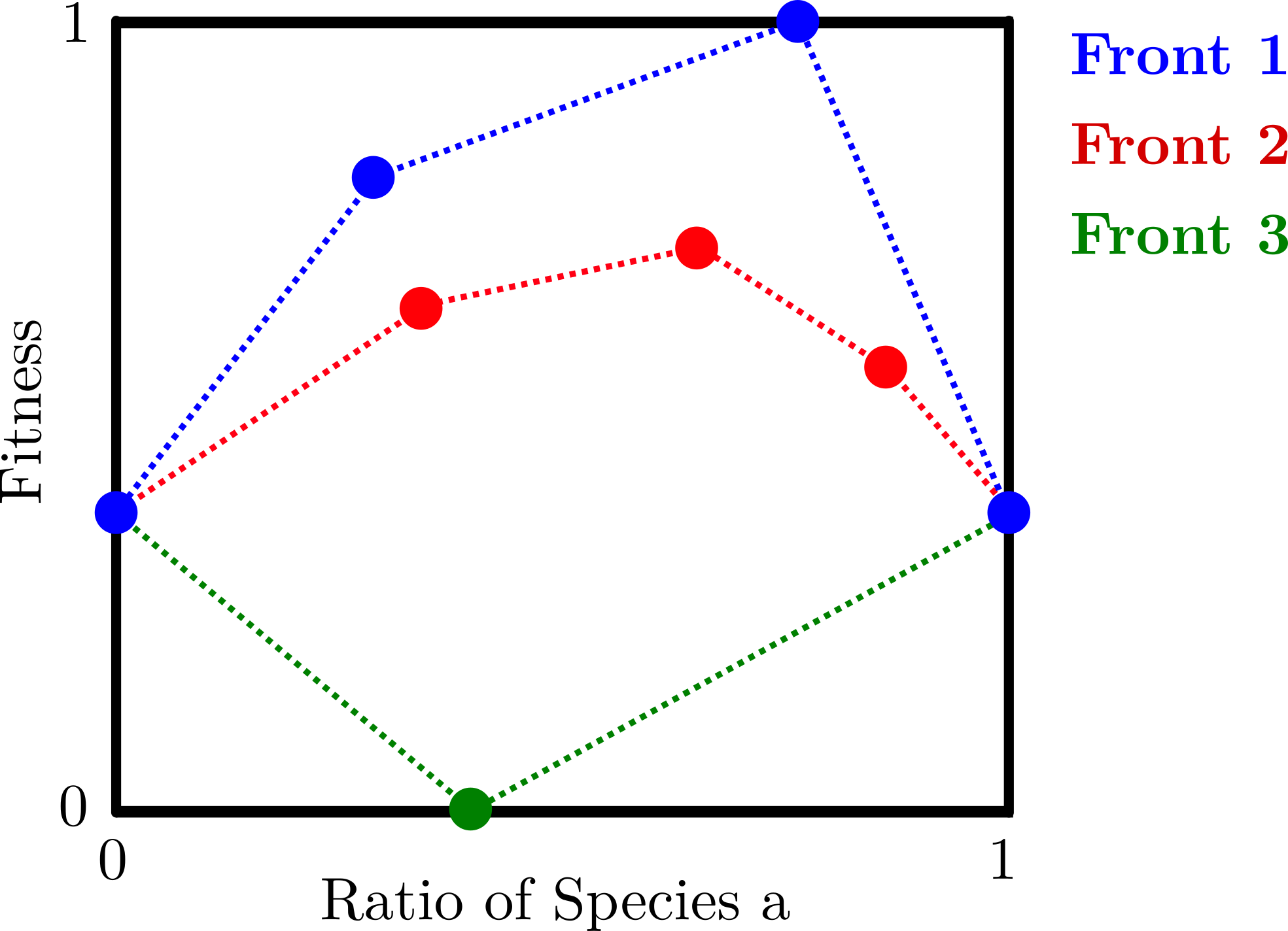}
  \caption{Illustrative example of front allocation for a two species CHGA, containing species $a$ and $b$, in fitness space. Members of front 1 are those on the Pareto CH and have a fitness $>$ the pure crystal structures. Front 2 is the Pareto CH of all members after removal of those on front 1 (including the pure cell points for the sake of hull evaluation though pure cells are always on front 1).}
  \label{fig:CH_front_allocation}
\end{figure}

A CH is not strictly a Pareto front as the standard `direction' of optimisation present in Pareto optimisation is ill defined for a CH. However, a minor modification to the NSGA approach allows it to be used for fitness evaluation in the CHGA. The CH of a given population can be seen as analogous to the Pareto optimal set or front 1; it is the `best known' set of thermodynamically stable structures where each member can be argued to be `equally as good' with respect to the CH. The `next best' members are front 2, those that exist on the Pareto optimal hull after removal of front 1, and so on. In the CHGA, fitness rather than binding enthalpy should be considered due to structural penalisation (\Secref{sec:convex_hull_in_fitness_space}). CH front allocation is illustrated in \Figref{fig:CH_front_allocation} and \Algorref{alg:CHGA_front_sorting}.

\begin{figure}[htbp]
\begin{algorithm}[H]
  \caption{Convex Hull Front Assignment}
  \label{alg:CHGA_front_sorting}
  \begin{algorithmic}

    \Input
    \Desc{$M$}{The set of population members.}
    \Desc{$m(\mathbf{R},\mathcal{F})\ \forall\ m \in M$}{Reduced species ratios $m(\mathbf{R})$ and fitness $m(\mathcal{F})$ for all $m \in M$.}
    \Desc{$f$}{The fitness of a member with 0 binding enthalpy (scaled appropriately).}
    \Desc{$S$}{The set of $(\mathbf{R},f)$ for all possible pure single species crystals.}
    \EndInput

    \Output
    \Desc{$m(P)\ \forall\ m \in M$}{Pareto front for all input members.}
    \EndOutput

    \vspace{1em}

    \For{$m \in M$} \Comment{Set all pure crystals in $M$ to be on front 1}
      \If {$m \in S$}
        \State $m(P) \gets 1$
      \Else
        \State $m(P) \gets -1$ \Comment{s.t. $m(P)<0$ means front unassigned for member}
      \EndIf
    \EndFor

    \State $C \gets 1$ \Comment{The current front being assigned}

    \While{$\exists\ m(P) < 0$} \Comment{Loop until all members have been assigned a front}

      \If{$\exists\ m \in M \text{ s.t. } m(\mathcal{F}) > f \text{ and } m(P) < 0$} \Comment{Do above and below fitness of $f$ separately}

        \State{$H\gets$ members on convex hull from $S \cup \{ m(\mathbf{R},\mathcal{F})\ \forall\ m \in M\ |\ m(\mathcal{F}) > f \text{ and } m(P)<0 \}$ }

      \Else

        \State{$H\gets$ members on convex hull from $S \cup \{ m(\mathbf{R},\mathcal{F})\ \forall\ m \in M\ |\ m(\mathcal{F}) \leq f \text{ and } m(P)<0 \}$ }

      \EndIf

      \For{$m \in H$ and $m \not\in S$} \Comment{Assign front to members on current hull}
        \State{$m(P) \gets C$}
      \EndFor

      \State{$C \gets C+1$}

    \EndWhile

  \end{algorithmic}
\end{algorithm}
\end{figure}


\subsection{Hull Hypervolume \& Convergence}

Convergence in CSP is when the method has found what it considers to be the optimal solution(s). Sub-optimal CSP methods converge slowly or do not find the true optimal solution(s) and the best CSP methods converge to the optimal solution(s) quickly. There are two main approaches to convergence in CSP, the first is to assume convergence has been achieved when the optimal structure(s) have not changed for some time (many generations in the case of a GA)
\cite{wang_calypso_2012,abraham_periodic_2006}. A second additional convergence parameter can be applied requiring known optimal final structure(s) to have been found multiple times \cite{lu_ab_2021}, though this is mainly applicable in cases where structural penalisation is not carried out.

Neither approach is appropriate for the CHGA, the latter as the CHGA utilises structural penalisation. Considering the first approach, it is possible to keep track of all population members on the hull, if there is any change in this set the hull is still improving and has not yet converged. However, this does not numerically measure how much a hull changes which would not, for example, allow for the comparison of convergent behavior of two different methods.

\begin{figure}[htbp]
    \centering
    \begin{subfigure}[t]{0.4\textwidth}
        \centering
        \includegraphics[width=\linewidth]{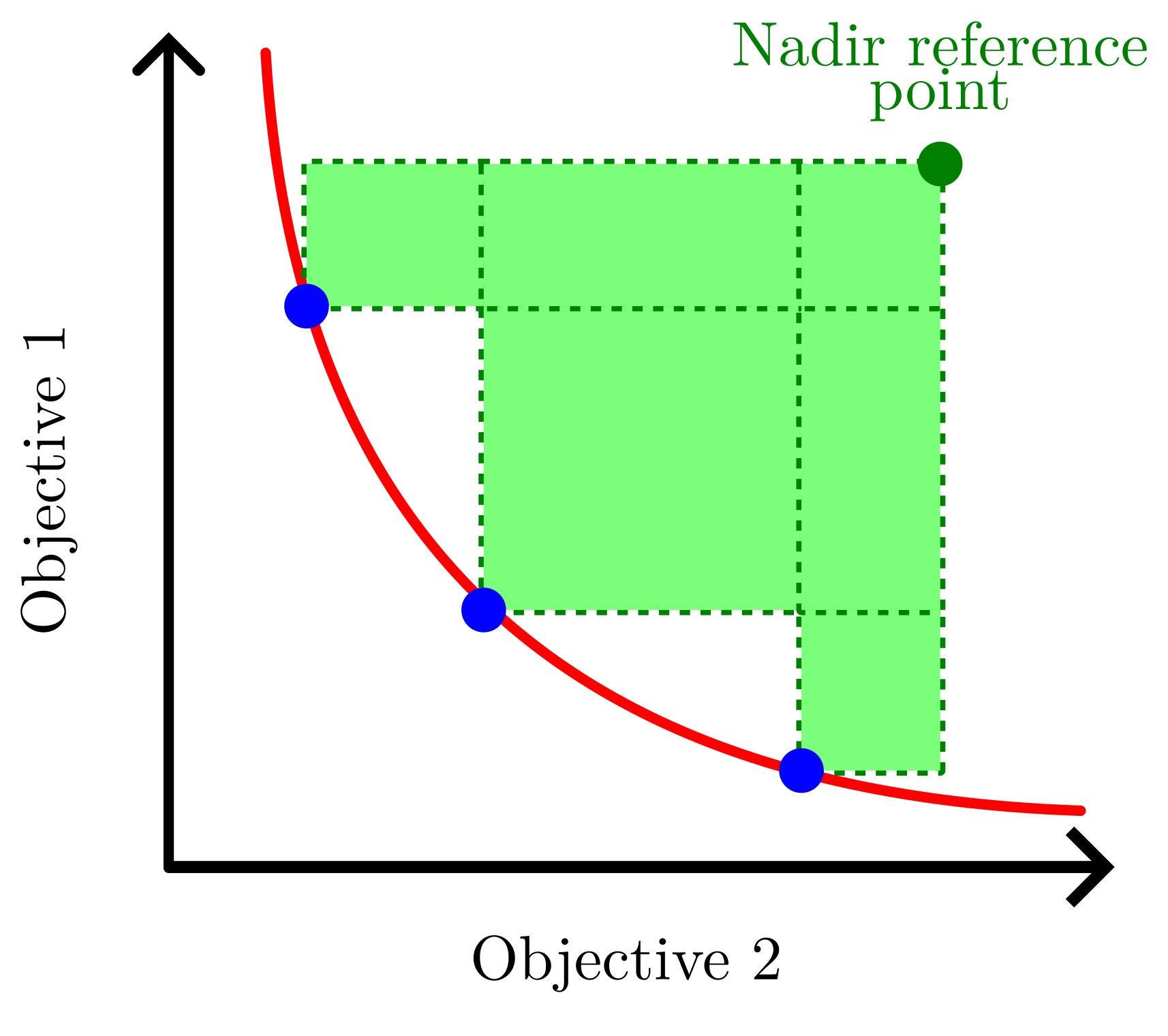}
        \caption{}
    \end{subfigure}
    \hfill
    \begin{subfigure}[t]{0.4\textwidth}
        \centering
        \includegraphics[width=\linewidth]{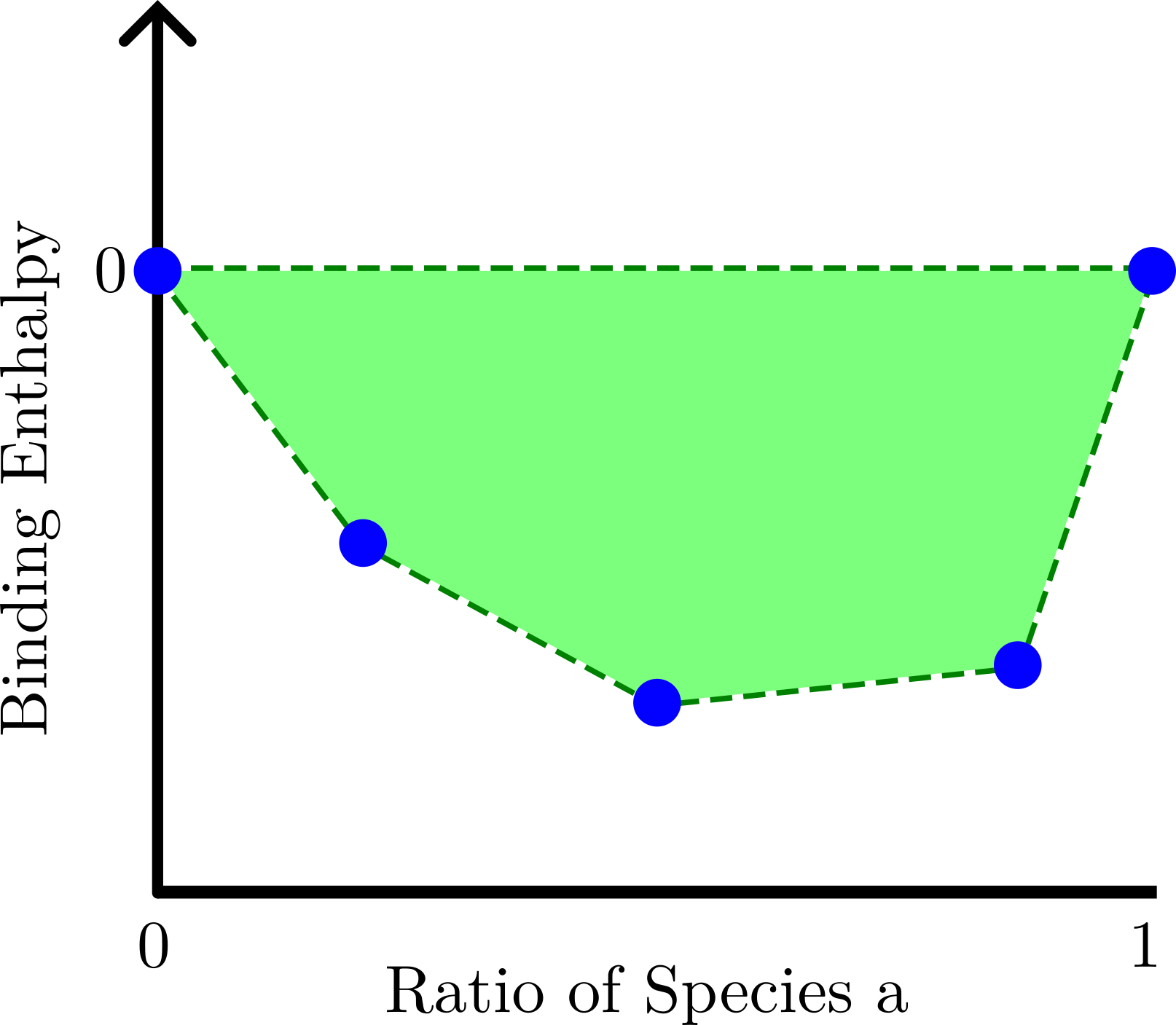}
        \caption{}
    \end{subfigure}

    \caption{Example of hypervolume calculation for a Pareto optimisation (a) \cite{fonseca_improved_2006,deb_toward_2010,garcia-najera_improved_2011} and in the CHGA (b), where the shaded area (in the 2D case) is the hypervolume. \SHH has a negative sign in (b) as it exists below 0 binding enthalpy.}
  \label{fig:hypervolume}
\end{figure}

A similar issue is encountered when trying to assess the convergence of a MOGA, with the $N$ dimensional hypervolume employed as a solution \cite{fonseca_improved_2006,deb_toward_2010,garcia-najera_improved_2011} (\Figref{fig:hypervolume}). However, as the CH is not a Pareto front the MOGA hypervolume approach is not directly applicable due to the Nadir reference point \cite{deb_toward_2010} not having an obvious position. However, the CHGA can use a hypervolume approach with a very minor modification.

As such, let a CH's \textit{hypervolume} be the contained space within the hull, for two species this is the area of the hull, for three species the volume, \textit{etc}. The hypervolume always has units of energy, as it is a product of a binding enthalpy and dimensionless species ratios. Then, let the \textit{signed hull hypervolume} (\SHH) be the hypervolume with the sign of the binding enthalpy of the multi-species structures on the hull (\Figref{fig:hypervolume}). The sign of the binding enthalpy for all multi-species cells on the hull must be the same as it is not possible for a convex hull to cross the edge joining all pure cells with 0 binding enthalpy; the signed hypervolume is negative for a hull containing thermodynamically stable structures.

The convergent behavior of \SHH can be used to test convergence `on the fly' to see if a run can be terminated due to convergence. \SHH can also be used to compare multiple CHGA runs that consider the same ionic species and trivially extended to a \textit{signed front hypervolume} (the signed hypervolume of a CH front with the addition of the pure cells).

\subsection{Niching}
\label{sec:niching}

Population diversity is an important consideration for a MOGA as (without some discouragement) MOGAs can evolve towards a Pareto optimal set that is tightly grouped on the Pareto front. The front method alone (\Secref{sec:pareto_hull_and_front}) does not discourage this tight grouping but it can be counteracted by \textit{niching}, a secondary fitness parameter generally based on how isolated a member is in parameter space. Niching discriminates only between members with identical primary fitness, \ie members of the same front.

\begin{figure}[htbp]
    \centering
    \begin{subfigure}[t]{0.4\textwidth}
        \centering
        \includegraphics[width=\linewidth]{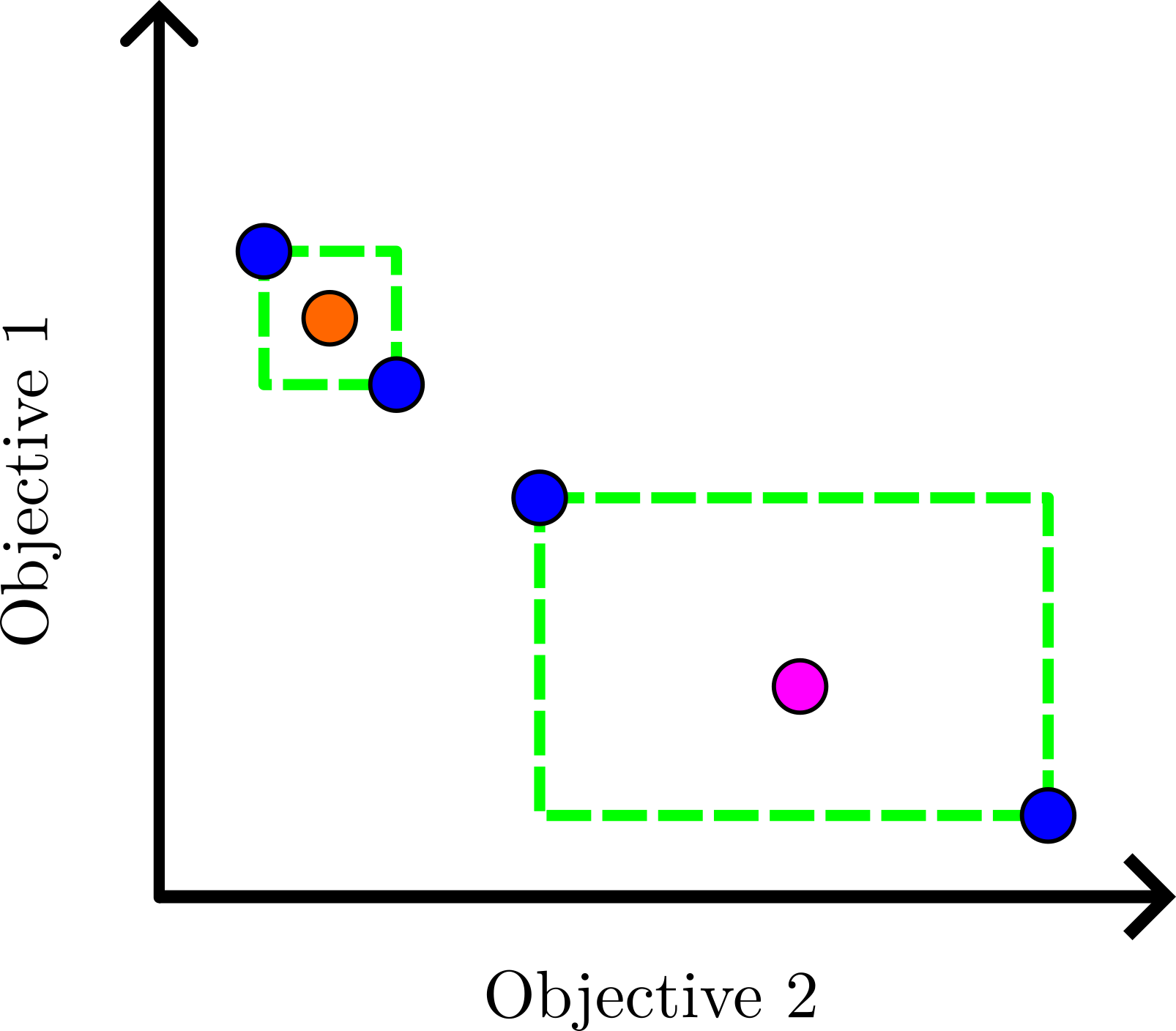}
        \caption{}
        \label{fig:niching_example_MOGA}
    \end{subfigure}
    \hfill
    \begin{subfigure}[t]{0.4\textwidth}
        \centering
        \includegraphics[width=\linewidth]{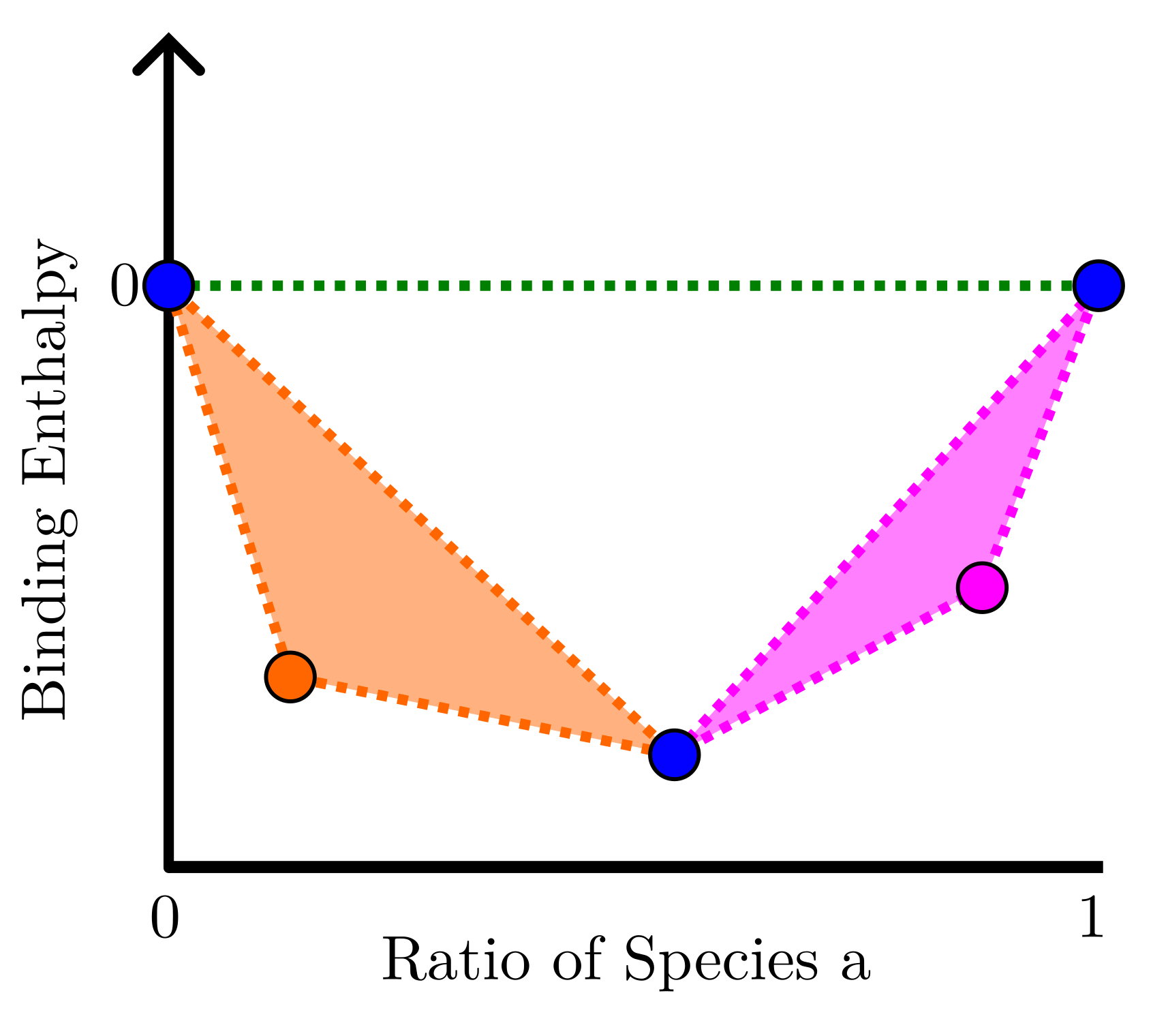}
        \caption{}
        \label{fig:niching_example_CHGA}
    \end{subfigure}

    \caption{Illustration of the niching method in NSGA-II (a) and CHGA (b). In both cases the orange point has a smaller (worse) niching value than the purple point. For NSGA-II this is as the rectangle with corners at the closest members has a smaller perimeter than the same construction around the purple member. For CHGA this is as the difference between the full hull hypervolume and the hypervolume of the hull under removal of the points (the highlighted areas) is larger for the purple point.}
  \label{fig:niching_example}
\end{figure}

Often the choice of niching method is a distinguishing factor between MOGA methods \cite{bosman_balance_2003}. For example, in NSGA-II niching is derived from the perimeter of the rectangle with corners defined by the closest members in parameter space (\Figref{fig:niching_example_MOGA}) \cite{deb_fast_2002}, with a more complex hyperplane approach used for $>2$ objective functions using NSGA-III \cite{deb_evolutionary_2014,jain_evolutionary_2014}. As niching is such an important part of the MOGAs that inspired the CHGA, consideration should be given to the application of niching in the CHGA.

\subsubsection{Does Niching Come `For Free' in the CHGA}

The space of thermodynamically stable structures is discontinuous with respect to a change in ionic ratio for the CHGA due in part to a constraint being applied to the maximum number of ions in a member (this is done for reasons of computational cost). Furthermore, the distribution of ionic ratios that it is possible to explore is unevenly distributed (see \Appref{sec:farey_discretisation_ionic_ratio}). As such, any difference in species ratio of two members on the CH implies two distinct structures. This is an important distinction to a standard MOGA, \ie proximity with respect to the parameter space of species ratios does not necessarily imply strong structural similarity.

This removes a concern that niching is designed to answer, \ie in the CHGA there is no reason to discourage members from being close to each other when they are on the same front. Members that are on the same front with different ionic ratios, no matter how close this ratio is, are not the same structure and therefore equally as interesting with respect to the CH. Therefore, the realities of the discrete space of species ratios could mean niching is `built in' to the CHGA front based fitness approach such that the increase in diversity provided by the niching could harm the efficiency of CHGA convergence.

\subsubsection{Hull Hypervolume Based Convex Hull Niching}

An increase in the dimensionality of a MOGA often leads to an increase in complexity of niching parameter, \eg the change in complexity and cost from NSGA-II to NSGA-III as dimensionality increases from 2 to $>2$ dimensions \cite{deb_fast_2002,deb_evolutionary_2014,jain_evolutionary_2014}. It is possible this is completely avoided by the discretisation of the ratios of the species in the convex hull parameter space meaning niching is not required in the case of the CHGA. However, to investigate if this is the case requires a CHGA niching approach.

\SHH could be argued to be the true optimisation parameter for the CHGA (a similar argument is sometimes made with respect to standard MOGAs \cite{hutchison_indicator-based_2004}). \SHH is a continuous parameter (due only to the binding enthalpy) and not all members on a front contribute equally to its \SHH. Therefore, a reasonable niching value is the change in \SHH of a front when a given member is removed (\Figref{fig:niching_example_CHGA} \& \Algorref{alg:CHGA_niching}), the larger the change in hull hypervolume of a member the `more important' it is to that hull. This should be carried out in fitness space such that the structural similarity penalisation is also considered.

\begin{figure}[htbp]
\begin{algorithm}[H]
  \caption{Convex Hull Niching}
  \label{alg:CHGA_niching}
  \begin{algorithmic}

    \Input
    \Desc{$M$}{The set of population members.}
    \Desc{$m(P)\ \forall\ m \in M$}{Pareto front for all input members where $m(P) \in \{ 1, \dots, \max(m(p))\}$.}
    \EndInput

    \Output
    \Desc{$m(\dshh)\ \forall\ m \in M$}{Niching values for all input members.}
    \EndOutput

    \vspace{1em}

    \For{$F=1,\dots,\max(M_{F})$}

      \State $P_{f} \gets \{ m \in M\ |\ m(F) = f \}$ \Comment{Set of all $m \in M$ on front $f$}
      \State $A \gets$ hypervolume of $P_{f}$ \Comment{Hypervolume of front $f$}

      \For{$m \in M$}

        \If{$f = m(F)$}

          \State $B \gets$ hypervolume of $P_{f}\cap m$ \Comment{Hypervolume of front excluding this member}
          \State $m(\dshh) \gets |A-B|$

        \EndIf

      \EndFor

    \EndFor

  \end{algorithmic}
\end{algorithm}
\end{figure}

The change in front hypervolume niching parameter ($\dshh$) depends on the front, therefore can only be used to differentiate members within a front, the larger \DSHH the better. Also, the increase in dimensionality of \DSHH could mirror the increase in complexity of MOGA niching values with an increase in number of objective functions.


\subsection{Member Selection in the CHGA}
\label{sec:member_selection}

Which members should be chosen for genetic operations, whether breeding new children or inter-generational survival, is based on member fitness. Roulette selection is one method routinely used for such selections (often used alongside other methods including rank and tournament selection) \cite{razali_genetic_2011,greenstein_determining_2023,abraham_improved_2008,curtis_gator_2018}. In roulette selection a population member is randomly selected with a weight proportional to the fitness parameter; roulette selection is more likely to choose a fit member, but can (with a lower probability) choose a lower fitness member.

\subsubsection{Breeding Selection}
\label{sec:breeding_selection}

Two distinct parents are chosen for breeding using roulette selection of the primary fitness characteristic, the front number. A member of the chosen front can either be randomly selected (assuming all members of each front are equally fit) or niching can be applied with a second roulette selection using \DSHH for the fitness weighting (with larger \DSHH more likely to be chosen). It is not clear from theoretical considerations along which of the two approaches is superior, so both will be tested.

\subsubsection{Inter-Generational Survival \& Elitism}
\label{sec:elitism}

When choosing which members from a population of parents and children should survive to be parents of the next generation (\ie using greedy selection) there are two main approaches, the first is some level of random (often roulette) selection (similar to breeding selection, \Secref{sec:breeding_selection}). However, some level of \textit{elitism} is often applied, where the next generation is selected (in part or entirely) from the fittest members of the previous generation. \Eg in NSGA-II the best fronts are added in full until a front is encountered that if added in full would cause an overflow with respect to population size, with the members of this front with the largest niching value surviving into the next generation \cite{deb_fast_2002}.

In GAs a \textit{semi-elitist} approach can be used; a proportion of the best members are automatically selected for survival (with the possible `immortality' of the fittest members reducing the risk of a genetic drift away from optimal solutions \cite{chang_wook_ahn_elitism-based_2003}), with the rest of the next generation selected with some random (often roulette) selection (to maintain genetic diversity) \cite{hutchison_semi-elitist_2006}. Semi-elitist approaches are specifically useful for MOGAs. This is because optimisation towards a Pareto optimal set may not occur at the same rate across the whole Pareto front; some parts of the front may be harder to reach than others. If purely elitist selection is used members that are sub-optimal but evolving towards unrepresented parts of the Pareto front may be discarded \cite{hutchison_semi-elitist_2006}.

In the case of the CHGA it is clear elitism should be applied at least to front 1, the best representation of the CH currently known, as this is the subject of optimisation and without maintaining front 1 the magnitude of the signed hull hypervolume would decrease. However, the exact level of \textit{front based elitism} to use (\ie how many of the best fronts to take in full into the next generation) is unclear. Generally $>1$ fronts would have to be considered to find all members `on or near the hull', but exactly how many is not obvious and may be system dependant. As such, the level of front based elitism was tested, with results in \Secref{sec:results_elitism}.


\subsection{From 2 to 3 Species CHGAs}

All of the approaches to front based fitness allocation and hypervolume based convergence/niching calculation outlined so far apply to any number of dimensions (\ie species) $\geq 2$. However, there is an extra subtlety that must be considered when using a CHGA for CSP with more than 2 species. For brevity this section considers a 3 species CHGA, but the arguments extend to any $N>2$ dimensions.

The main issue is that the \textit{`edges'} of a three species CH; the three two species hulls that bound the edges of the valid ionic ratios in parameter space. If the two-species `edges' are under-populated then three species members could be erroneously ascribed to the CH when they actually lie within it.

However, to always breed a pair of two species children both parents must either be pure cells or consist of only the same two species. Breeding a two species child from one or two three species parents is also possible, though the parents would have to fulfill specific constraints. Therefore, assuming the population is not heavily dominated by a set of members containing the same two species, it is more likely when using front based roulette selection to breed three species children in a three species CHGA. This leads to under explored two species `edges' which can drastically effect the CH.

A simple way to rectify this is to perform a CHGA for each of the edges individually, performing three two species CHGA runs to individually find the two species parts of the three species hull. Two species `edges' can then be used as the initial population for a three species CHGA. This increases the number of required CHGA runs, though the increase in efficiency to the final three species CHGA may offset the cost of the `extra' three 2D CHGAs. The increase in number of calculations is also non-linear with number of dimensions, \eg to initialise a four species hull in this way requires six 2D and four 3D hulls. Also, an increase in cost would not be surprising, as rarely does the cost of an optimisation problem increase linearly with an increase in dimensionality.


\subsection{The Initial Population \& High Symmetry Structures in the CHGA}
\label{sec:high_symmetry_structures}

The first step in any genetic algorithm is to generate an initial population on which to apply evolutionary operations. In the CHGA the ground state \textit{pure} single species crystal cells, the enthalpy of which is needed to calculate binding enthalpies, are included in the initial population. Pure crystal structures can provide structural motifs that are important genetic information when breeding members on the hull (especially in the dopant dominated regions of the hull). These cells must be known ground state structures, which can be found with a standard GA or from a structural database as the ground states for single species crystals are generally well explored.

The remainder of the initial population is constructed from generated high symmetry cells, evenly distributed across the three dimensional space groups. In general at RTP it is more likely that structures in space groups $>1$ will lie at global minima (with some notable exceptions like glasses \textit{etc}). As such, initial exploration of the PES is restricted to high-symmetry structures, which are generally more `genetically useful' than completely random ones. The advantage of high symmetry searching can be seen from its successful use in other CSP methods \cite{wang_calypso_2012,pickard_ab_2011,lee_crystal_2021}.

Further, from a purely genetic standpoint, symmetry is more likely to evolve in a population even without a symmetry based selection pressure \cite{johnston_symmetry_2022}. This means a genetic approach to CSP will likely favor symmetric crystals. As there is no universal optimal CSP algorithm (\ie there is `no free lunch' \cite{wolpert_no_1997}) \cite{pickard_ab_2011,falls_xtalopt_2021}, different CSP algorithms will not be equally optimal for all CSP problems. Therefore, genetic approaches are likely best suited to optimisation of structures where symmetry is important, which is most crystal structures of interest.

The predisposition of genetic operators towards symmetric systems is inherent to the flow of hereditary information in evolution \cite{johnston_symmetry_2022}. As such, one may expect an initial high symmetry population to be closer to what would be expected after some generations (regardless as to what evolutionary pressure is applied). In effect the high symmetry initial population leans into the strengths of a genetic approach.

From these arguments it follows that a large initial high symmetry population would be useful. However, an extremely large initial population would increase the cost required before the first application of genetic processes. Due to the complete lack of heredity in the high symmetry cells they can be seeded into child populations each generation. This helps maintain symmetry diversity in the populations whilst allowing the use of hypervolume convergence to terminate the CHGA after only the required number of cell evaluations, avoiding the question of how many initial high symmetry cells are required.

In the CHGA high symmetry cells are generated by allocating ions to Wyckoff sites for a randomly selected space group within a cell. Lattice angles $\alpha,\beta,\gamma$ and the ratios of lattice vector lengths $|\mathbf{a}|$ to $|\mathbf{b}|$ and $|\mathbf{b}|$ to $|\mathbf{c}|$ are randomly generated within the restrictions of the space group. $|\mathbf{a}|,|\mathbf{b}|,|\mathbf{c}|$ are then assigned (maintaining their ratios) based on a V\'egard's Law like approach, using the pure cell densities rather than pure cell lattice vectors. The density rather than lattice vector length (as is usual for V\'egard's Law) due to the differing lattice types required by different space groups.

Due to stochastic elements of cell generation and breeding, ionic positions of generated or bred cells can contain ions that are unrealistically close together. Cells containing such an `overlap' of ions are re-generated, re-bred or re-mutated. However, it is not immediately obvious what the minimum ionic separation should be, as it is both species-dependent (the normal separation for \ce{H2} is much less than than the minimum physical separation for a hypothetical \ce{Pb2} dimer) and environment-dependent (ionic separation at high pressure will be smaller than for RTP). Minimum ionic separation is defined as a scalar multiple of the species-dependent pseudopotential radius (the scalar allows for environment dependant modification, \eg reduction for high pressure systems).


\section{Testing Methods \& Machine Learned Potentials}
\label{sec:testing_methods_and_ML}

The number of geometry optimisations in a single CHGA testing run can be up to $\OO{10^{4}}$ (\eg $\OO{10^{2}}$ members over $\OO{10^{2}}$ generations). Due to the stochastic elements of the CHGA a random high symmetry initial population could for example `get lucky' and contain the optimal set, which would not be much of a test of the genetic operations. This means when testing a CHGA multiple runs (differing only by random seed) are required, increasing the number of geometry optimisations for each test to $\OO{10^{5}}$.

Though \textit{ab initio} methods (such as DFT) would be ideal for performing the cell relaxations, the cost for the required number of calculations makes such methods impractical for testing. This is especially true as the results for the repeated runs will likely converge to the same optimal population. In the development and testing of previous CSP methods empirical classical potentials have been used to address this problem \cite{abraham_improved_2008,falls_xtalopt_2021,wang_calypso_2012,lyakhov_new_2013}. Empirical potentials often require careful parameterisation such that they maintain some physical relevance and oversimplified potentials can lead to very rapid convergence of genetic methods meaning they can be unsuitable as a proxy potentials for the more realistic DFT potential.

Machine learned (ML) potentials have gained popularity in recent years as their cost in prediction of (at least) the energy, stress and strain of a given structure has a cost comparable to an empirical potential with their results (given the correct architecture and training) giving a closer reflection of the true PES \cite{riebesell_matbench_2024}. It is important to note that ML potentials are less reliable than \textit{ab initio} methods and require training on some representative dataset which is relatively costly, both in acquisition of the dataset and performing the requisite training.

Fortunately, many available ML methods provide a pre-trained model which have been tested and cataloged in the Matbench Discovery leaderboard \cite{riebesell_matbench_2024}. The pre-trained CHGNet potential \cite{deng_chgnet_2023}, which has been trained on $\sim 1.5 \times 10^{6}$ structural relaxations covering 89 elements from the Materials Project database \cite{deng_chgnet_2023,jain_commentary_2013,riebesell_matbench_2024}, was chosen for CHGA testing due to a combination of evaluation speed and accuracy (as measured by Matbench Discovery \cite{riebesell_matbench_2024}).

Although ML potentials are reliable for many structures, care should be taken when using any method as a `black box', especially alongside a CSP method. However, we have found that the pre-trained CHGNet potential offers sufficient complexity to serve as a good proxy for the DFT potential without requiring additional training. The lack of additional system-specific training does impact the accuracy of the ML-potential, for example, there could exist an extreme minima in a PES, possibly the result of extremely distorted structures (understandably) not being present in the training dataset (see \Appref{app:CHGNEt_close_example} for an example). If this is the case the CSP method will find these minima, which are correct with respect to the given PES, but far from physically relevant. This is, however, a sign that the CSP method is working correctly, and fully physical results may be acquired either by further training of the ML potential or the use of \textit{ab initio} methods.

Extra training based on systems of interest using the pre-trained model as a starting point would be required to refine the ML potential. This training requires extra cost however, so is avoided for the purposes of testing CHGA. To account for any possible erroneous behavior, any relaxed structure with ionic overlap (as high pressure systems are not considered this is defined as overlap of spheres of radius 0.8 times CASTEP C19 pseudopotential \cite{CASTEP,OTFG_C9} radius centered on each ion) are discarded during CHGA testing.

In brief, whilst pre-trained ML potentials may be expected to fail to reproduce the \textit{ab initio} PES with full agreement, especially for extremely distorted cells, they are of sufficient complexity to test the CSP algorithms. Furthermore, within their range of validity they are sufficiently close to the \textit{ab initio} potential to provide qualitatively useful results.


\section{Results}
\label{sec:results}

The CH of lithium silicide was used to test the CHGA, chosen in part as this system is of interest due to its possible application to future battery technology \cite{valencia-jaime_novel_2016,wang_reversible_2007}. As such, comprehensive CSP studies are available; previous CSP results of note (used for comparison in this section) are from AIRSS \cite{morris_thermodynamically_2014}, the GASP GA \cite{tipton_structures_2013,GASP-Python} and minima hopping \cite{valencia-jaime_novel_2016}, summarised alongside known experimental structures in \cite{valencia-jaime_novel_2016}.

Structural calculations were performed using the pre-trained CHGNet potential \cite{deng_chgnet_2023} (see \Secref{sec:testing_methods_and_ML}) to predict the energy, stress and forces on each structure alongside LBFGS \cite{BFGS,LBFGS} as implemented within CASTEP \cite{CASTEP} to perform structural relaxation.

\begin{figure}[htbp]
    \centering
    \begin{subfigure}[t]{0.48\textwidth}
        \centering
        \includegraphics[width=\linewidth]{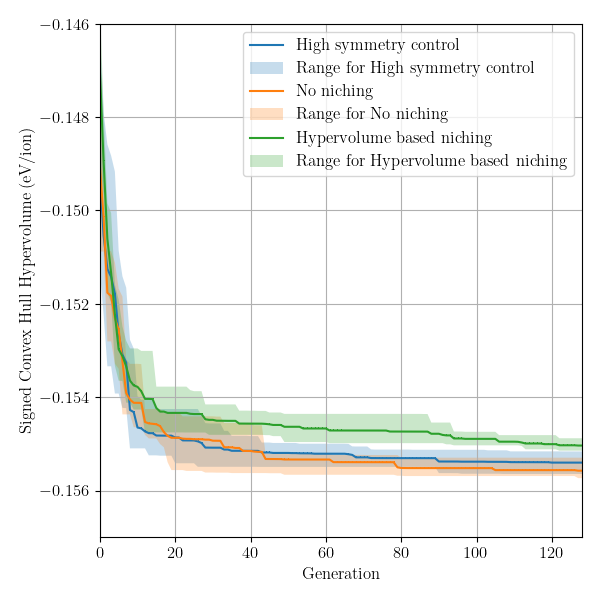}
        \caption{CHGA convergence test of niching method. The high symmetry only control is shown against no niching and hypervolume based niching. Non-control runs are fully elitist and contain 50/50 high symmetry and bred children per generation.}
        \label{fig:LiSi_conv_niching}
    \end{subfigure}
    \hfill
    \begin{subfigure}[t]{0.48\textwidth}
        \centering
        \includegraphics[width=\linewidth]{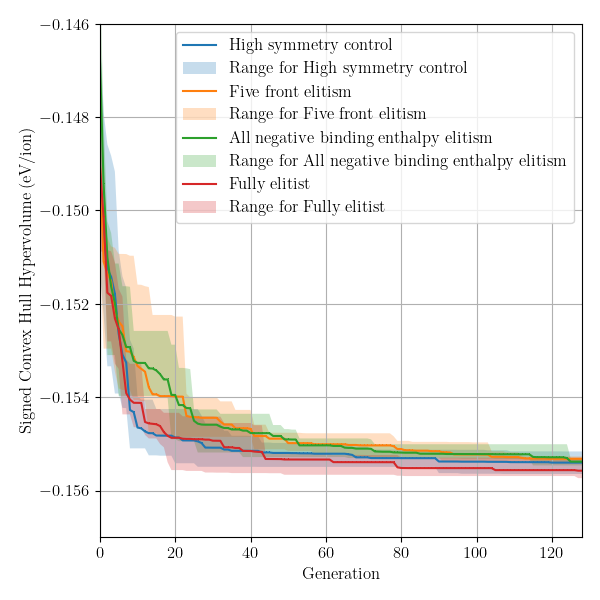}
        \caption{CHGA convergence test of elitism methods. The high symmetry control is shown against elitism with the five best fronts, elitism taking all fronts containing members fitter than pure cells and full elitism. Non-control runs do not use niching and contain 50/50 high symmetry and bred children per generation.}
        \label{fig:LiSi_conv_elitism}
    \end{subfigure}
    
    \vspace{0.5cm}
    
    \begin{subfigure}[t]{0.48\textwidth}
        \centering
        \includegraphics[width=\linewidth]{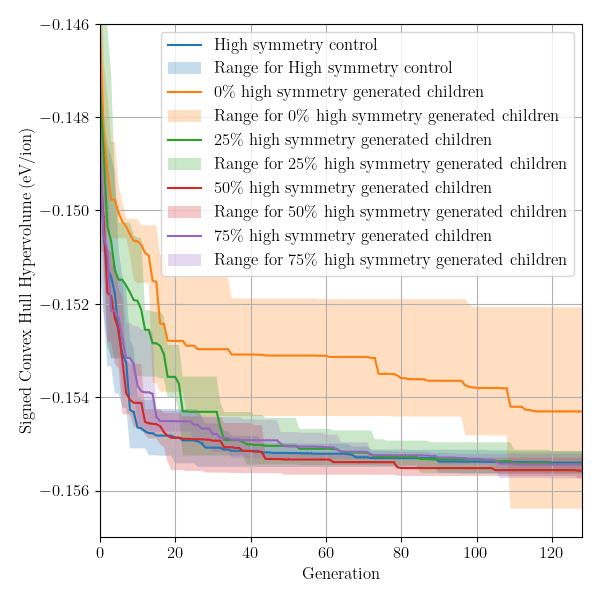}
        \caption{CHGA convergence test of high symmetry seeding. The high symmetry control is shown against 0\%, 25\%, 50\% and 75\% of children being generated as high symmetry structures. Non-control runs do not use niching and are fully elitist..}
        \label{fig:LiSi_conv_high_sym}
    \end{subfigure}
    \caption{Selected convergence tests of the CHGA for the LiSi system. The mean hypervolume and the range of hypervolumes of the hulls over 4 CHGA runs (differing only by random seed) is shown for each convergence test. For clarity the niching (a), elitism (b) and high symmetry seeding (c) is considered separately.}
    \label{fig:LiSi_convergence_tests}
\end{figure}

Convergence tests for CHGA algorithm parameters are shown in \Figref{fig:LiSi_convergence_tests} and discussed individually in the following subsections. To account for the stochastic elements of the CHGA all convergence tests are averaged over four runs (each differing only by random seed). A complete hull from using the parameters from the optimally converged CHGA run is shown in \Figref{fig:LiSi_CH}, where all members from the 4 runs with binding enthalpy $\leq 0$ and within 0.05 eV/ion of the hull are considered.

Convergence tests were carried out by measuring hull hypervolume over 128 generations containing 128 members restricted to contain 12 to 48 ions. A CHGA containing only randomly generated high symmetry cells as children (\ie without breeding operations) was used as a control for each test and runs were ran for a fixed number of generations rather than being terminated on reaching convergence. Convergence tests were applied to niching (Sections \ref{sec:niching} \& \ref{sec:member_selection}), elitism (\Secref{sec:elitism}) and the number of randomly generated high symmetry cells seeded into each child population (\Secref{sec:member_selection}), with results of convergence tests given in \Figref{fig:LiSi_convergence_tests}.

\subsection{Convergence Tests for Niching}

Niching is found to be sub-optimal for a LiSi CHGA (\Figref{fig:LiSi_conv_niching}) as the final \SHH is above that of all runs of both that without niching and the control. However, when considering the generated hull (\Figref{fig:LiSi_CH}) it should be noted that the majority of members on or near the hull are concentrated along a small portion of the CH (most in the range 60\% to 85\% Li). As niching encourages an `even spread' of bred members along the hull then, for this system, it discourages breeding from the small section of the hull where optimal members lie. However, it is impossible to know the distribution of members on the hull \textit{a priori} for an un-explored system, so it cannot be known without any \textit{a priori} knowledge of the phase space if niching will help or hinder convergence for a two species CHGA.

Such issues are less likely when considering three species convex hulls, where the density of members on the hull can differ between the three two species `edges'. If such a discrepancy exists niching is required in order to encourage breeding of population members across all parameter space (rather than biased towards the single high density CH `edge'). However, if the population is approximately uniformly distributed such that the niching values for all members are similar it will have little effect. Therefore, as niching adds negligible cost and will either benefit or have little effect on a three species hull its use is recommended in such cases. This will be tested explicitly in future work.

\subsection{Convergence Tests for Elitism}
\label{sec:results_elitism}

Completely front based elitism (the next generation made only from the best fronts) was tested alongside elitism only for the five best fronts and all members with fitness better than the pure cells (with other members of the next generation selected with roulette selection), as discussed in \Secref{sec:elitism}.

A fully elitist approach was found to be more optimal for the CHGA than other forms of elitism (\Figref{fig:LiSi_conv_elitism}), converging both faster and to a lower \SHH. This differs from other CSP GA methods that utilise partial elitism as optimising the convex hull directly necessitates the survival of various stoichiometries, \eg the members of front 1 of the CHGA will always contain more than one stoichiometry (\ie at least the single species cells).

\subsection{Convergence Tests for High Symmetry Cell Seeding}

The effect on convergence of seeding of randomly generated high symmetry cells into child populations (\Secref{sec:high_symmetry_structures}) was investigated by running the CHGA for LiSi whilst seeding 0\%, 25\%, 50\%, 75\% and 100\% (the high symmetry only control) high symmetry cells into the child population each generation.

\begin{figure}[htbp]
  \centering
  \includegraphics[width=0.98\textwidth]{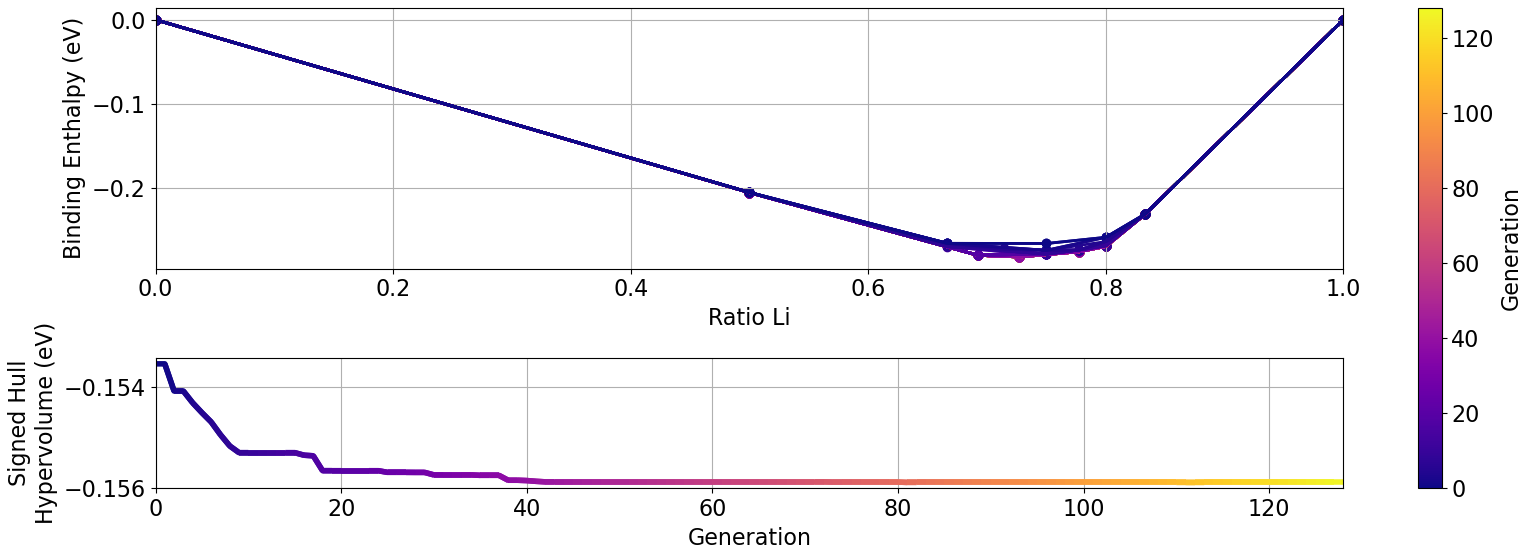}
  \caption{Analysis of the convergence of the CH from the most optimal CHGA LiSi run, using no niching, full elitism and a 50/50 mix of high symmetry and bred children. The top panel shows the hull plot at each generation with the hypervolume shown on the bottom panel. It can be seen that convergence is achieved rapidly, with all members on the hull found by generation 38.}
  \label{fig:LiSi_convergence_closer_look}
\end{figure}

A 50/50 mix of high symmetry generated cells and bred children was found to be optimal (\Figref{fig:LiSi_conv_high_sym}), with all structures on the hull found by generation 38 (\Figref{fig:LiSi_convergence_closer_look}). The this is the optimal method from those tested, as high symmetry seeding not only helps maintain symmetry based diversity, but (as can be seen from the LiSi hull in \Figref{fig:LiSi_CH}) different points on (or near) the hull are represented by bred and high symmetry members respectively. This suggests a high symmetry random structure search may be optimal for some structures whilst genetic methods optimal for others, which is no surprise when considering both the success of AIRSS \cite{morris_thermodynamically_2014,cen_exploring_2023,ferreira_search_2023,pickard_high-pressure_2006,lu_ab_2021} and the no free lunch theorem \cite{wolpert_no_1997}. Furthermore, high symmetry seeding is complimentary to the genetic methods as all optimal members, whether generated as a  high symmetry structure or bred, may be considered for parenthood by the genetic methods.

\subsection{The LiSi Convex Hull}

\begin{figure}[htbp]
  
    \centering
    \begin{subfigure}[t]{0.95\textwidth}
        \centering
        \includegraphics[width=\linewidth]{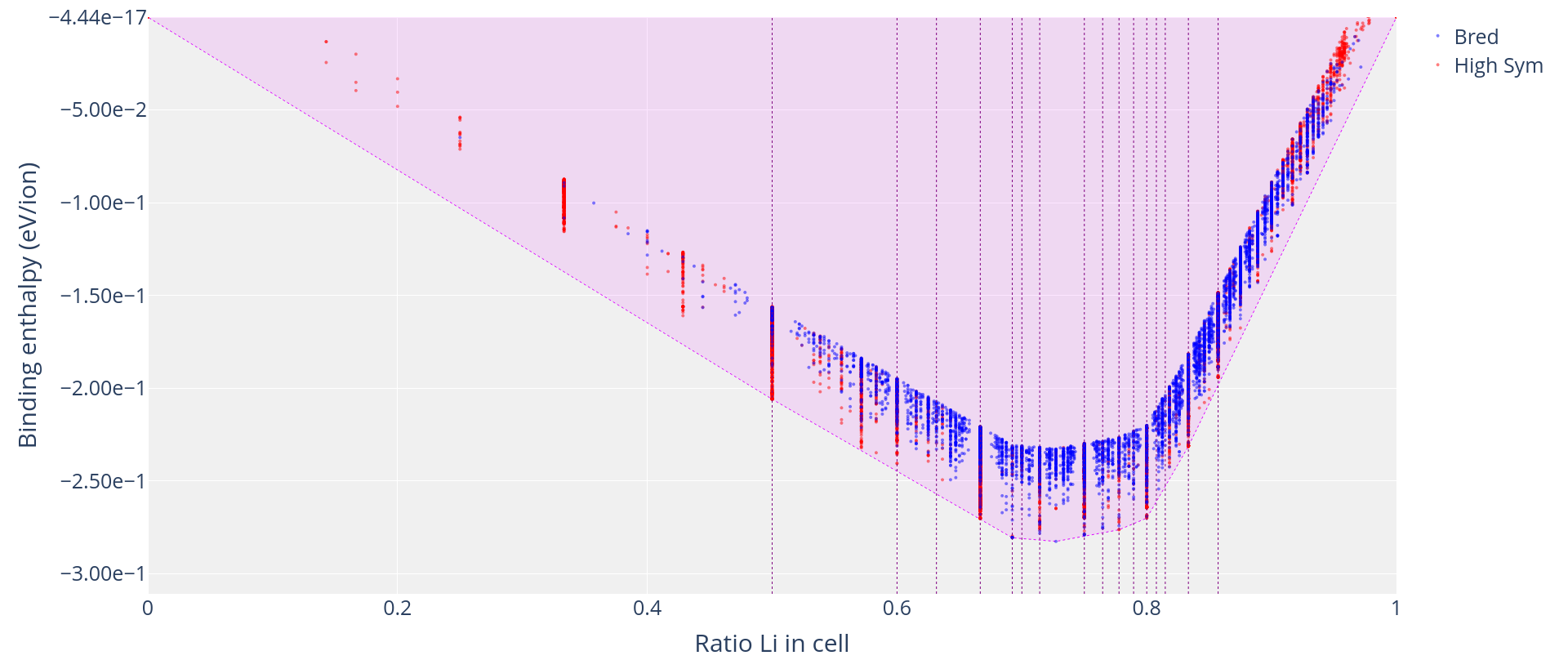}
    \end{subfigure}
    
    \vspace{0.5cm}
    
    \begin{subfigure}[t]{0.95\textwidth}
        \centering
        \includegraphics[width=\linewidth]{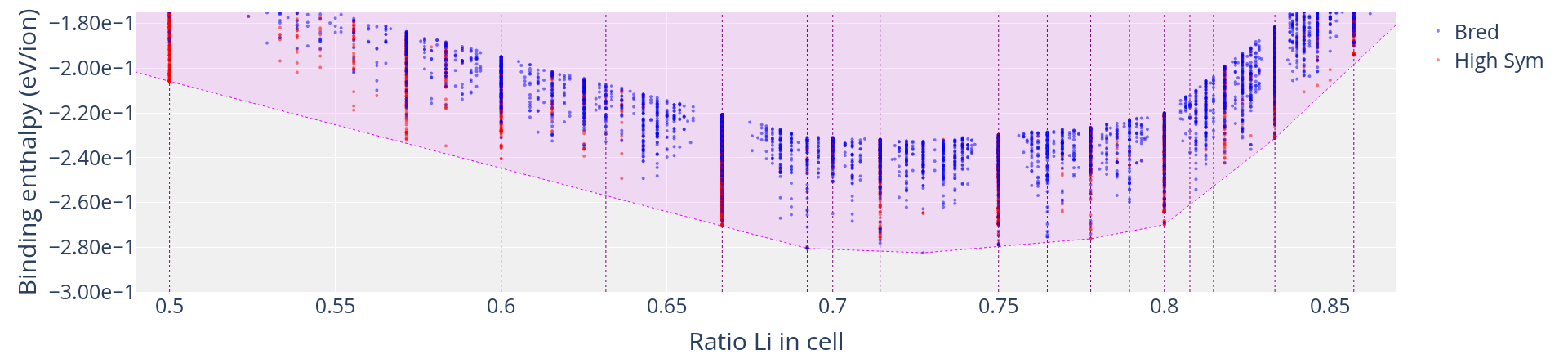}
    \end{subfigure}

    \caption{The LiSi convex hull as found by the CHGA using a population size of 128 where every structure on the hull was discovered by generation 38. Bred members are plot in blue and high symmetry members in red. Vertical lines represent ratios where structures are reported to exist \cite{morris_thermodynamically_2014,tipton_structures_2013,valencia-jaime_novel_2016}. Top panel shows the full hull and bottom the species ratios containing the most members on or near the hull.}
    \label{fig:LiSi_CH}
\end{figure}

The hull generated from the 50/50 high symmetry and bred child population with full elitism and no niching is shown in \Figref{fig:LiSi_CH}. Members within  0.05 eV/ion ($k_{b}T$ for $T \approx 580$) of the hull are defined to be `near the hull', with only these members shown in \Figref{fig:LiSi_CH}. Vertical lines are at ionic ratios where known structures exist on or near the previously reported hulls (\ce{LiSi}, \ce{Li3Si2}, \ce{Li12Si7}, \ce{Li2Si}, \ce{Li9Si4}, \ce{Li7Si3}, \ce{Li5Si2}\ ce{Li3Si}, \ce{Li13Si4}, \ce{Li7Si2}, \ce{Li15Si4}, \ce{Li21Si5}, \ce{Li4Si}, \ce{Li22Si5}, \ce{Li5Si}, and \ce{Li6Si}) \cite{valencia-jaime_novel_2016,morris_thermodynamically_2014,tipton_structures_2013}. The CHGA found structures on or near the hull for all of the known stoichiometries. Binding enthalpies do differ from the literature \cite{valencia-jaime_novel_2016} though this is due probably different DFT parameterisation to the CHGNet training data.

More structures can be seen on the hull (with many more near it) in \Figref{fig:LiSi_CH} than have been previously reported \cite{valencia-jaime_novel_2016}. No claims are made to the physicality of these structures without further DFT validation. However, the CHGA has replicating known results and presenting new structures that require further study shows the power of this approach.

In general, high symmetry cells are closer to the hull than bred members for the LiSi hull in \Figref{fig:LiSi_CH}. As a small element of noise $\OO{0.01\text{ \AA}}$ is applied to break symmetry in the initial generated high symmetry structure before local optimisation, it would be expected if a valid structure is found by a high symmetry search it would be close to the local minima. Conversely, as bred members are subjected to non-symmetrised stochastic processes they are likely to be initialised further from equilibrium than high symmetry cells. This implies it is likely geometry optimisation is more important for an identical final bred and high symmetry cell on the hull; bred cells are less likely to have converged geometries, so may be further away from the hull than optimal high symmetry members. Therefore, for a high throughput CHGA caution should be applies when deciding how close to the hull a member should be (especially if it is a result of breeding) for consideration for later low-throughput refinement and analysis.


\FloatBarrier

\section{Conclusion}

A genetic approach to crystal structure prediction of a convex hull was presented. This approach builds on the methods in the CASTEP GA \cite{abraham_periodic_2006,abraham_improved_2008,higgins_simultaneous_2019} using a novel algorithm inspired by multi-objective optimisation. Specifically, the convex hull was approached as the object of optimisation utilising a convex hull front based fitness and niching parameter. The convex hull hypervolume was utilised as a convergence parameter to compare the convergent behaviour of the introduced methods.

A LiSi convex hull was considered using the pre-trained CHGNet machine learned potential \cite{deng_chgnet_2023} for cell evaluations. Convergence analysis on this system showed that niching was not critical in the case of this two species system, full elitism should be used for the CHGA alongside a hybrid approach using a 50/50 split of bred and randomly generated children.

Using CHGNet the convex hull genetic algorithm was shown to be able to discover all known stable stoichiometries on the two species LiSi convex hull that are reported in the literature from AIRSS, GASP GA, minima hopping and experimental approaches \cite{morris_thermodynamically_2014,tipton_structures_2013,valencia-jaime_novel_2016}. The CHGA also found novel candidate structures in the pre-trained CHGNet PES. Through this testing the CHGA was shown to be an efficient and powerful tool for the automated discovery of novel thermodynamically stable multi-species crystal structures.

Future work will involve testing the conjecture that niching is more important in the case of three (or more) species convex hulls than they are with two species hulls. Also, a test of other chemical systems where the hull is not dominated by one region of species ratios is an additional step with respect to generalised efficiency gains for niching for two species hulls. Finally, as the CHGA will run with any number of species $\geq 2$ studies with three or more species looking at both explored and novel structures is a natural next test of the CHGA.


\section*{ACKNOWLEDGEMENTS}
We are grateful for computational support from the University of York with the Viking cluster, which is a high performance compute facility provided by the University of York. We are also grateful for computational support from the UK Materials and Molecular Modelling Hub (which is partially funded by EPSRC EP/P020194) and from the UK national high performance computing service, ARCHER2. Access for both of these was obtained via the UKCP consortium and funded by EPSRC grant ref EP/X035891/1. The authors also acknowledge EPSRC grant EP/V047779/1 for financial support.


\bibliography{main}

\clearpage
\begin{appendices}

\section{Distribution of Stoichiometries in the CHGA}
\label{sec:farey_discretisation_ionic_ratio}

Due to cost limitations the number of allowed ions in a cell in the CHGA is restricted to within a user defined range. Unsurprisingly, this limitation excludes stoichiometries that require more than the maximum number of ions to create. For example, if an \ce{LuNH} convex hull is being considered and the maximum number of ions in a cell is limited to 10 then a \ce{Lu4N2H5} stoichiometry would be unobtainable (and therefore the CH partially incomplete \cite{ferreira_search_2023}).

However, what is less obvious is that the allowed ionic ratios are not evenly distributed. This is easiest to consider with two species, let $M,N$ be the minimum and maximum number of total ions allowed in a cell respectively. Then recall that a Farey sequence of order $N>0$ is the set of rational numbers in $[0,1]$ that can be expressed as fractions with denominator $\leq n$ \cite[\S 5.4]{apostol_modular_1976}, explicitly
\begin{equation}
  F_{N}
  :=
  \left\{ \frac{a}{b} \in [0,1] \text{ s.t. } 0 < b \leq N \text{ and } 0 \leq a \leq b \right\}
  \quad .
\end{equation}
The ionic ratios that are possible to explore are in
\begin{equation}
  F_{N}
  \supset
  F_{N}^{M}
  :=
  \left\{ \frac{a}{b} \in [0,1] \text{ s.t. } 0 < M < N \text{ , } M \leq b \leq N \text{ and } 0 \leq a \leq b \right\}
  \quad .
\end{equation}
The Farey sequence is not evenly spaced on $[0,1]$, which implies the same is true for $F_{N}^{M}$, which implies uneven discretisation of the ionic ratios. \Eg there is generally a large space around ionic ratio of 0.5 that cannot be explored unless $n$ is large. This argument extends to $>2$ ionic species.

A consequence of discretisation of $F_{N}^{M}$ is the increased probability of the CHGA exploring certain ionic ratios. This follows from a combinatorial argument; given some reasonably sized $N$ there are more ways of constructing certain ionic ratios than others. \Eg there are more ways of constructing an ionic ratio of 1/2 than 1/4 (within the bounds of the allowed number of ions). Specifically, the difficulty in exploration of an ionic ratio scales with the required denominator.

Therefore, more easily constructed ionic ratios are more likely to be explored by the CHGA. With this in mind, $N$ must be large enough to allow for reasonable exploration of the parameter space; if there are only a few ways to construct a given ionic ratio it is less likely this ratio will be fully explored (when compared to a less difficult to construct ratio).

Practically, this means $N$ must be large enough for all reasonable stoichiometries to be explored and this choice of $N$ makes exploration of certain stoichiometries less likely. A heuristic arrived at by experimentation is that one may expect the CHGA to possibly struggle with cells that require $\gtrapprox N/2$ ions in order to create the stoichiometry. \Eg for \ce{Lu4N2H5} one should likely set $N \gtrapprox 22$.


\section{CHGNet Energy for Compressed Diamond Si Cell}
\label{app:CHGNEt_close_example}

\begin{figure}[htbp]
  \centering
  \begin{subfigure}{.5\textwidth}
    \centering
    \includegraphics[width=.9\linewidth]{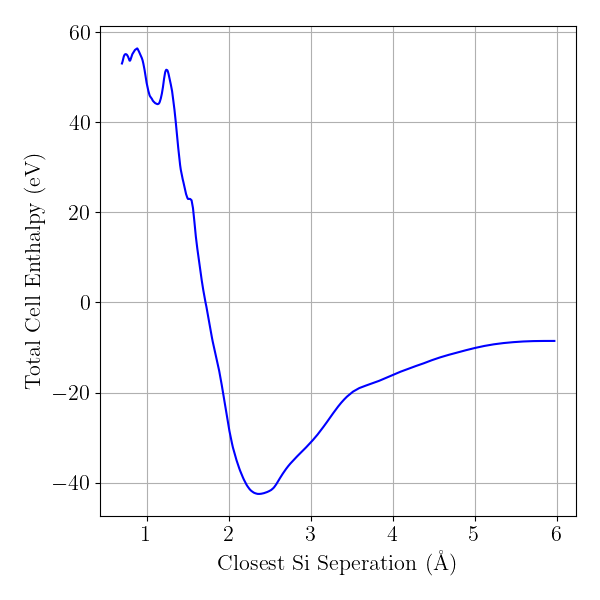}
  \end{subfigure}%
  \begin{subfigure}{.5\textwidth}
    \centering
    \includegraphics[width=.9\linewidth]{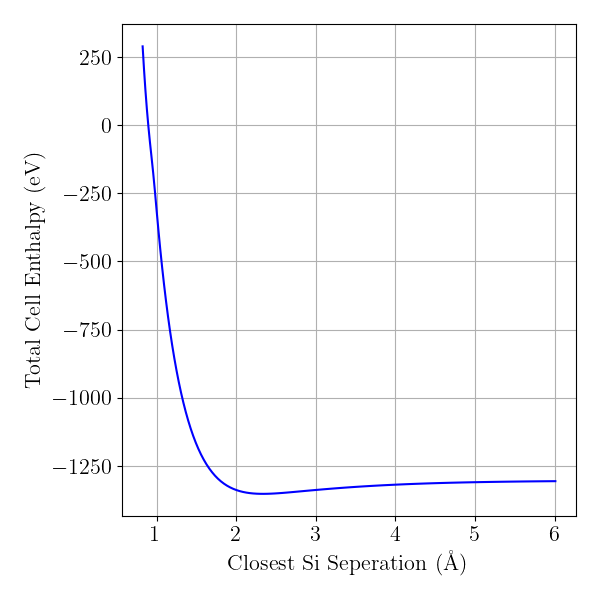}
  \end{subfigure}
  \caption{The total cell enthalpy of an 8 ion primitive Si cell under a change of cell volume calculated with pre-trained CHGNet (left) and DFT (right).}
  \label{fig:CHGNet_Si_seperation}
\end{figure}

The pre-trained CHGNet potential was used to calculate the enthalpy of an eight ion conventional diamond structure silicon cell, where $|\mathbf{a}|=|\mathbf{b}|=|\mathbf{c}|$, for a varying lattice vector lengths. This is shown in \Figref{fig:CHGNet_Si_seperation} alongside a similar DFT calculation for the same structure. DFT calculations were carried out with CASTEP \cite{DFT-HK,DFT-KS,MPgrid,RMP-Payne,USP,CASTEP,PZ-LDA} using the local density approximation (LDA) exchange correlation functional \cite{PZ-LDA}, CASTEP C19 pseudopotential \cite{USP,OTFG_C9}, cut off energy of 300 eV and k-point spacing of 0.0428 $\text{\AA}^{-1}$.

The unusual behavior occurs below and around the pseudopotential core radii for silicon ($\sim 0.95$ \AA\ for LDA C19). Due to the challenge of DFT calculations with ionic separations much lower than pseudopotential radii it is unlikely that structures with such Si ionic separations are well represented within the pre-training dataset. Note, it was not the LDA C19 ultra-soft pseudopotentials that were used in the pre-trained DFT calculations \cite{deng_chgnet_2023}, but this argument assumes a comparable pseudopotential radius.

\begin{figure}[htbp]
  \centering
  \begin{subfigure}{.5\textwidth}
    \centering
    \includegraphics[width=.9\linewidth]{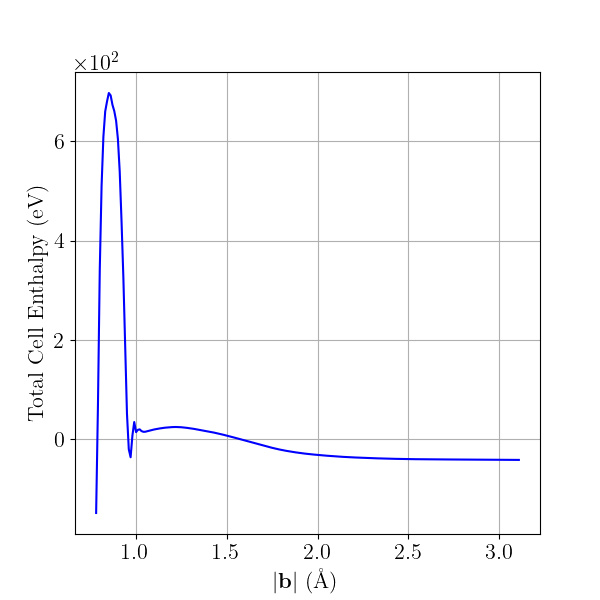}
  \end{subfigure}%
  \begin{subfigure}{.5\textwidth}
    \centering
    \includegraphics[width=.9\linewidth]{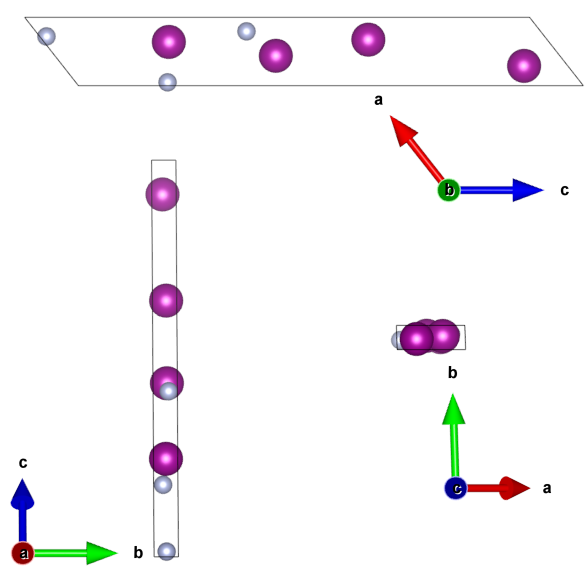}
  \end{subfigure}
  \caption{The change in energy calculated with pre-trained CHGNet (left) of the MnN cell (right, clockwise from top along $\mathbf{b}$ then $\mathbf{c}$ then $\mathbf{a}$) as $|\mathbf{b}|$ is varied (keeping $|\mathbf{a}|=2.788663\ \text{\AA}$ and $|\mathbf{c}|=16.166506$ \AA).}
  \label{fig:odd_cell_vary_b}
\end{figure}

During initial CHGA testing similar behaviour was noted with pre-trained CHGNet with, for example, the cell in \Figref{fig:odd_cell_vary_b} (generated during an early CHGA testing run). This unusual cell's strange behavior is a result of the very close ionic separations; it has a very (and erroneously) low enthalpy predicted by pre-trained CHGNet. However, due to the spacing of the ions the slicing breeding means all children of this structure are very likely to inherit the closely separated ions, which leads to more of the same problem. Though this cell is actually low on the pre-trained CHGNet PES, it is not a reasonable test with respect to an actual PES.

This is unequivocally not a criticism of CHGNet (pre-trained or otherwise), and similar behavior can be seen in other pre-trained ML models trained on the same dataset such as MACE \cite{Batatia2022mace,Batatia2022Design}. It is instead likely this arises from such nonphysical systems being under or un-represented in the training dataset, due in part to it being inadvisable to have a lot of pseudopotential core overlap in a DFT calculation. However, CSP methods like the CHGA will exploit anything they can to find minima in the PES, even if these minima are far from physically relevant.

\end{appendices}

\end{document}